\documentclass[aps,prd,showpacs,twocolumn]{revtex4}
\usepackage{amssymb}
\usepackage{amsfonts}
\usepackage{amsmath}
\usepackage{graphicx}
\usepackage{color}
\usepackage[dvips]{epsfig}
\usepackage[dvips]{graphicx}
\usepackage{float}

\begin{document}

\title{Magnetic flux inversion in Charged BPS vortices in a
Lorentz-violating Maxwell-Higgs framework}
\author{R. Casana$^{1}$, M. M. Ferreira Jr.$^{1},$ E. da Hora$^{1,3},$ C.
Miller$^{1,2}$}
\affiliation{$^{1}$Departamento de F\'{\i}sica, Universidade Federal do Maranh\~{a}o,
65085-580, S\~{a}o Lu\'{\i}s, Maranh\~{a}o, Brazil.}
\affiliation{$^{2}$Instituto de F\'{\i}sica Te\'orica, UNESP-Universidade Estadual
Paulista}
\affiliation{R. Dr. Bento T. Ferraz 271 - Bl. II, 01140-070, S\~ao Paulo, SP, Brasil}
\affiliation{$^{3}$Departamento de F\'{\i}sica, Universidade Federal da Para\'{\i}ba,
58051-900, Jo\~{a}o Pessoa, Para\'{\i}ba, Brazil.}

\begin{abstract}
We demonstrate for the first the existence of electrically charged
BPS vortices in a Maxwell-Higgs model supplemented with a parity-odd
Lorentz-violating (LV) structure belonging to the CPT-even gauge sector of
the standard model extension and a fourth order potential (in the absence of
the Chern-Simons term). The modified first order BPS equations provide
charged vortex configurations endowed with some interesting features:
localized and controllable spatial thickness, integer flux quantization,
electric field inversion and  {localized} magnetic flux reversion.
 {This model could possibly be applied on condensed matter systems
which support charged vortices carrying integer quantized magnetic flux,
endowed with localized flipping of the magnetic flux.}
\end{abstract}

\pacs{11.10.Lm,11.27.+d,12.60.-i}
\maketitle

Since the seminal works by Abrikosov \cite{ANO1} and Nielsen-Olesen \cite%
{ANO2} showing the existence of electrically neutral vortices in type-II
superconducting systems and in field theory models, respectively, these
nonperturbative solutions have been a theoretical issue of enduring
interest. In the beginning 90's, vortices solutions were studied in the
context of planar theories including the Chern-Simons term \cite{CS} which
turned possible the attaining of electrically charged vortices \cite{CSV}
also supporting BPS (Bogomol'nyi, Prasad, Sommerfeld) solutions \cite{BPS},
related with the physics of anyons and the fractional quantum Hall effect
\cite{Ezawa}. In addition, charged BPS vortices also were found in the
Maxwell-Chern-Simons model \cite{MCS}. Further studies were performed
involving nonminimal coupling \cite{CSV1} and new developments \cite%
{CSV2,Bolog}. Generalized Chern-Simons vortex solutions were recently
examined in the presence of noncanonical kinetic terms (high order
derivative terms) \cite{Hora1} usually defined in the context of k-field
theories \cite{kfield,Cosmology,GC}. These k-defects present a compact-like
support \cite{Compact1}, an issue of great interest currently \cite{Compact2}%
. Lately, in the context of effective field theories, it has also been
demonstrated the existence of charged BPS vortices in a generalized
Maxwell-Chern-Simons-Higgs model \cite{Cadu-last}.

Lorentz-violating (LV) theories have been under attention in the latest
years. The general theoretical framework for studying Lorentz-violation
effects is the standard model extension (SME) \cite{Colladay} which
encompasses Lorentz-violating terms in all sectors of the minimal standard
model. In particular, the abelian gauge sector of this model is composed of
two sectors, a CPT-odd and a CPT-even one. The CPT-even part is described by
nineteen parameters enclosed in the rank 4 tensor, $\left( K_{F}\right)
^{\mu \alpha \nu \beta },$ endowed with a double null trace and the
symmetries of the Riemann tensor \cite{KM1}, being investigated in many
respects \cite{Even}. Lorentz violation was also considered in connection
with the formation of topological defects \cite{Monopole1,Monopole2}, with
particular interest in the Higgs sector \cite{Defects}. Recently, it has
been investigated the formation of stable uncharged vortices in the context
of the nonbirefringent Lorentz-violating and CPT-even Maxwell-Higgs
electrodynamics, also including LV terms in the Abelian Higgs sector of the
SME, with new interesting results \cite{Carlisson1}.

Up the moment it is known that abelian charged vortices are only defined in
models endowed with the Chern-Simons term. This remains valid even in the
context of highly nonlinear models, such as the Born-Infield electrodynamics
\cite{Moreno}. \ In this letter we report for the first time the existence
of abelian charged BPS vortices in a Maxwell-Higgs electrodynamics deprived
of the Chern-Simons term and endowed with CPT-even LV terms. This
achievement is ascribed to the CPT-even electrodynamics of the SME, whose
parity-odd coefficients entwine the electric and magnetic sectors \cite%
{Baileyx,Cadu2} in analogy to what happens in the models containing the
Chern-Simons term. The BPS solutions are attained by considering a
particular fourth-order potential, and can be interpreted as vortex
solutions in a dielectric medium \cite{Dielectric}. The charged vortex
solutions are localized, having spatial thickness controlled by the LV
parameter, and presenting localized magnetic flux and electric field
reversion in some radial region. This phenomenon could be of interest in
condensed matter physics, mainly in connection with superconductivity,
particularly with two-components superconducting systems \cite{PRL1}.

\section{The theoretical framework}

The theoretical environment in which our investigations are developed is a
modified Maxwell-Higgs model defined by the following Lagrangian density
\begin{align}
\mathcal{L}_{1+3}& =-\frac{1}{4}F_{\mu \nu }F^{\mu \nu }-\frac{1}{2}\kappa
^{\mu \nu }F_{\mu \rho }F_{\nu }^{\text{ \ }\rho }+\left\vert D_{\mu }\phi
\right\vert ^{2}  \notag \\[0.2cm]
& +\frac{1}{2}\left( 1+\kappa _{00}\right) \partial _{\mu }\Psi \partial
^{\mu }\Psi +\frac{1}{2}\kappa ^{\mu \nu }\partial _{\mu }\Psi \partial
_{\nu }\Psi   \label{L2} \\[0.2cm]
& -e^{2}\Psi ^{2}\left\vert \phi \right\vert ^{2}-U\left( \left\vert \phi
\right\vert ,\Psi \right) ,  \notag
\end{align}%
containing a convenient potential $U\left( \left\vert \phi \right\vert ,\Psi
\right) $. The two first terms in (\ref{L2}) define the nonbirefringent and
CPT-even electrodynamics of the SME, whose nine LV nonbirefringent
parameters are enclosed in the symmetric and traceless tensor,\ $\kappa
^{\mu \nu },$ defined as%
\begin{equation}
\kappa ^{\mu \nu }=\left( K_{F}\right) _{\text{ \ \ \ \ \ \ }\alpha }^{\mu
\alpha \nu },
\end{equation}%
where $\left( K_{F}\right) _{\text{ \ \ }}^{\mu \alpha \nu \beta }$\ is the
CPT-even gauge sector of the SME, whose properties were much investigated
since 2001 \cite{KM1,Even}. The Higgs field, $\phi $, is coupled to the
gauge sector by the covariant derivative, $D_{\mu }\phi =\partial _{\mu
}\phi -ieA_{\mu }\phi $. The neutral scalar field, $\Psi $, plays the role
of an auxiliary scalar field, and is analogous to the one that appears in
the planar Maxwell-Chern-Simons models endowed with charged BPS vortex
configurations \cite{MCS,Bolog}. Note that the Lorentz-violating tensor, $%
\kappa ^{\mu \nu },$ also modifies the kinetic term of the neutral field.
The potential in Eq. (\ref{pot}) which assures the attainment of BPS first
order equations is defined by
\begin{equation}
U\left( \left\vert \phi \right\vert ,\Psi \right) =\frac{\left(
ev^{2}-e\left\vert \phi \right\vert ^{2}-\epsilon _{ij}\kappa _{0i}\partial
_{j}{\Psi }\right) ^{2}}{2\left( 1-s\right) }.  \label{pot}
\end{equation}%
observe that it is a fourth-order one. Here, $v$ is the vacuum expectation
value of the Higgs field, while $s=\mbox{tr}\left( \kappa _{ij}\right) $.
The potential $U$ possesses a nonsymmetric minimum, $\Psi =0$, $\left\vert
\phi \right\vert =v$, which is responsible by providing topological charged
vortex configurations. We now analyze static solutions for the model
projected in the $xy-$plane, where it recovers the structure of a $(1+2)-$%
dimensional theory. \ Remembering that in $(1+2)-$D it holds $%
F_{ij}=\epsilon _{ij}B,$ $F_{0i}=E^{i}$ (the magnetic field becomes a
scalar), we write the stationary Gauss's law of the planar model related to (%
\ref{L2}) by
\begin{equation}
L_{ij}\partial _{i}\partial _{j}A_{0}{}+\epsilon _{ij}\kappa _{0i}\partial
_{j}B=2e^{2}A_{0}\left\vert \phi \right\vert ^{2},  \label{Gauss_1}
\end{equation}%
where
\begin{equation}
L_{ij}=\left( 1+\kappa _{00}\right) \delta _{ij}-\kappa _{ij},
\end{equation}%
\ carries on the CPT-even and parity-even LV parameters.\ We note that is
the parity-odd parameter $\kappa _{0i}$ that couples the electric and
magnetic sectors \cite{Baileyx,Cadu2}, making possible the existence of
charged vortex configurations which are attained even in absence of the
Chern-Simons term. Otherwise, for $\kappa _{0i}=0$, the temporal gauge $%
A_{0}=0$ solves the Gauss law, yielding compactlike uncharged vortex
solutions \cite{Carlisson1}.

\section{BPS construction}

In this section, we focus our attention on the development of a BPS
framework \cite{BPS} consistent with the second order differential equations
obtained from the $(1+2)-$dimensional version of the Lagrangian (\ref{L2}).
We begin writing the energy density $E$\ in the stationary regime\ as
\begin{align}
\mathcal{E}& =\frac{1}{2}\left( 1-s\right) B^{2}+U\left( \left\vert \phi
\right\vert ,\Psi \right) +\left\vert D_{j}\phi \right\vert ^{2}  \notag \\%
[0.1cm]
& +\frac{1}{2}L_{ij}\left( \partial _{i}A_{0}\right) \left( \partial
_{j}A_{0}\right) +\frac{1}{2}L_{ij}\left( \partial _{i}\Psi \right) \left(
\partial _{j}{\Psi }\right)   \label{energy_0} \\[0.2cm]
& +e^{2}A_{0}^{2}\left\vert \phi \right\vert ^{2}+e^{2}\Psi ^{2}\left\vert
\phi \right\vert ^{2}.  \notag
\end{align}

\ In order to attain the first order differential equations, we first impose
the following condition\ on the neutral field\ $\Psi $:%
\begin{equation}
\Psi =\mp A_{0},  \label{bpsCC-1}
\end{equation}%
which is similar to the ones appearing in the BPS vortex configurations of
the Maxwell-Chern-Simons model \cite{CSV2,Bolog}. \ By substituting (\ref%
{bpsCC-1}) in\ Eq. (\ref{energy_0}), we attain%
\begin{align}
\mathcal{E}& =\frac{1}{2}\left( 1-s\right) B^{2}+\frac{\left(
ev^{2}-e\left\vert \phi \right\vert ^{2}\pm \epsilon _{ij}\kappa
_{0i}\partial _{j}A_{0}\right) ^{2}}{2\left( 1-s\right) }  \notag \\[-0.2cm]
& \\
& +\left\vert D_{j}\phi \right\vert ^{2}+L_{ij}\left( \partial
_{i}A_{0}\right) \left( \partial _{j}A_{0}\right) +2e^{2}A_{0}^{2}\left\vert
\phi \right\vert ^{2}.  \notag
\end{align}%
After converting the two first terms in quadratic form and by using the
identity,%
\begin{equation}
\left\vert D_{j}\phi \right\vert ^{2}=\left\vert D_{\pm }\phi \right\vert
^{2}\pm e\left\vert \phi \right\vert ^{2}B\pm \frac{1}{2}\epsilon
_{ab}\partial _{a}J_{b},
\end{equation}%
where $J_{b}$\ is the spatial component of the current $J^{\mu }=i\left[
\phi \left( {D}^{\mu }\phi \right) ^{\ast }-\phi ^{\ast }{D}^{\mu }\phi %
\right] $, the energy density takes on the form%
\begin{align}
\mathcal{E}& =\frac{1}{2}\left( 1-s\right) \left[ B\mp \frac{%
ev^{2}-e\left\vert \phi \right\vert ^{2}\pm \epsilon _{ij}\kappa
_{0i}\partial _{j}A_{0}}{\left( 1-s\right) }\right] ^{2}  \notag \\
& +\left\vert D_{\pm }\phi \right\vert ^{2}\pm ev^{2}B\pm \frac{1}{2}%
\epsilon _{ab}\partial _{a}J_{b} \\[0.2cm]
& +L_{ij}\left( \partial _{i}A_{0}\right) \left( \partial _{j}A_{0}\right)
+B\epsilon _{ij}\kappa _{0i}\partial _{j}A_{0}+2e^{2}A_{0}^{2}\left\vert
\phi \right\vert ^{2}.  \notag
\end{align}%
Now, we use the Gauss's law to transform the last three terms in a total
derivative, rewriting the energy density as
\begin{align}
\mathcal{E}& =\frac{1}{2}\left( 1-s\right) \left[ B\mp \frac{%
ev^{2}-e\left\vert \phi \right\vert ^{2}\pm \epsilon _{ij}\kappa
_{0i}\partial _{j}A_{0}}{\left( 1-s\right) }\right] ^{2}  \notag \\
&  \label{E_den} \\
& +\left\vert D_{\pm }\phi \right\vert ^{2}\pm ev^{2}B+\partial _{a}\mathcal{%
J}_{a},  \notag
\end{align}%
where
\begin{equation}
\mathcal{J}_{a}=\pm \frac{1}{2}\epsilon _{ab}J_{b}+L_{ab}A_{0}\partial
_{b}A_{0}+\epsilon _{ba}\kappa _{0b}BA_{0}.
\end{equation}%
This energy density (\ref{E_den}) is minimized by requiring that the squared
terms be null, establishing the two BPS conditions of the model:
\begin{align}
& \displaystyle{\left\vert D_{\pm }\phi \right\vert =0~,}~\ \
\label{BPS_cart1} \\[0.3cm]
& \displaystyle{B=\frac{\pm \left( ev^{2}-e\left\vert \phi \right\vert
^{2}\right) +\epsilon _{ij}\kappa _{0i}\partial _{j}A_{0}}{\left( 1-s\right)
},}  \label{BPS_cart2}
\end{align}%
Under these BPS conditions, we attain the BPS energy density,
\begin{equation}
\mathcal{E}_{BPS}=\pm ev^{2}B+\partial _{a}\mathcal{J}_{a},  \label{den_1}
\end{equation}%
implying the total BPS energy,%
\begin{equation}
E_{BPS}={\pm }ev^{2}\int d^{2}rB={\pm }ev^{2}\Phi ,  \label{den_2}
\end{equation}%
which is proportional to the magnetic flux.\ It is worthwhile to note that
the second term in (\ref{den_1}) does not contribute to the total BPS
energy, once the fields go to zero at infinity.

\section{Charged vortex configurations}

Specifically, we look for radially symmetric solutions using the standard
static vortex Ansatz%
\begin{equation}
A_{\theta }=-\frac{a\left( r\right) -n}{er},~\phi =vg\left( r\right)
e^{in\theta },~A_{0}=\omega (r),  \label{ansatz}
\end{equation}%
that allows to write the magnetic field as
\begin{equation}
B=-\frac{a^{\prime }}{er}.  \label{cb1}
\end{equation}%
The scalar functions $a\left( r\right) ,$\ $g\left( r\right) $\ and $\omega
\left( r\right) $\ are regular at $r=0$\ and $r=\infty ,$\ satisfying the
appropriate boundary conditions:%
\begin{align}
g\left( 0\right) & =0,\;a\left( 0\right) =n,~\omega ^{\prime }\left(
0\right) =cte,  \label{bc00} \\[0.3cm]
g\left( \infty \right) & =1,\,a\left( \infty \right) =0,\,\omega \left(
\infty \right) =0,  \label{bc01}
\end{align}%
with $n$\ being the winding number of the vortex solution. The boundary
conditions above will be demonstrated explicitly in the remain of the
manuscript.

We now introduce the dimensionless variable $t=evr$\ and implement the
following changes: \
\begin{eqnarray}
&\displaystyle{g\left( r\right) \rightarrow \bar{g}\left( t\right) ,\
a\left( r\right) \rightarrow \bar{a}\left( t\right) ,\omega \left( r\right)
\rightarrow v\bar{\omega}\left( t\right) ,}&  \notag \\[-0.2cm]
&& \\
&\displaystyle{B\rightarrow ev^{2}\bar{B}\left( t\right) ,~{\mathcal{E}}%
\rightarrow v^{2}\bar{\mathcal{E}}\left( t\right) .}&  \notag
\end{eqnarray}%
Thereby, the BPS equations (\ref{BPS_cart1},\ref{BPS_cart2}) and the Gauss's
law (\ref{Gauss_1}) are rewritten in a dimensionless form%
\begin{eqnarray}
&\displaystyle{\bar{g}^{\prime }=\pm \frac{\bar{a}\bar{g}}{t}\,,}&
\label{eqt2} \\[0.25cm]
&\displaystyle{\bar{B}=-\frac{\bar{a}^{\prime }}{t}=\frac{\pm \left(
1-g^{2}\right) -\kappa \bar{\omega}^{\prime }}{\left( 1-s\right) }\,,}&
\label{eqt3} \\[0.3cm]
&\displaystyle{\left( 1+\lambda _{r}\right) \frac{\left( t\bar{\omega}%
^{\prime }\right) ^{\prime }}{t}-\kappa \frac{\left( t\bar{B}\right)
^{\prime }}{t}-{{2\bar{g}^{2}\bar{\omega}=0}}\,,}&  \label{eqt4}
\end{eqnarray}%
where $s=\mbox{tr}\left( \kappa _{ij}\right) =\kappa _{rr}+\kappa _{\theta
\theta }$\ and we have defined $\kappa =\kappa _{0\theta }$\ and $\lambda
_{r}=\kappa _{00}-\kappa _{rr}$. Also, the signal $+$\ corresponds to $n>0$\
and $-$\ to $n<0$. We can also observe from Eqs. (\ref{eqt2},\ref{eqt4})\
that under the change $\kappa \rightarrow -\kappa $, the solutions go as $%
\bar{g}\rightarrow \bar{g}~,\ \bar{a}\rightarrow \bar{a}~,~\bar{\omega}%
\rightarrow -\bar{\omega}$.

We now discuss the magnetic flux quantization. We first rewrite the BPS
energy density (\ref{den_1}) in terms of the ansatz (\ref{ansatz}),\ that
is,
\begin{equation}
{\bar{\mathcal{E}}\,}_{BPS}={\pm \bar{B}\pm \frac{\left( \bar{a}\bar{g}%
^{2}\right) ^{\prime }}{t}+\left( 1+\lambda _{r}\right) \frac{\left( t\bar{%
\omega}\bar{\omega}^{\prime }\right) ^{\prime }}{t}-\kappa \frac{\left( t%
\bar{\omega}\bar{B}\right) ^{\prime }}{t}}.  \label{ebps0}
\end{equation}%
The first term is the magnetic field whose integration under boundary
conditions (\ref{bc00},\ref{bc01}) gives the magnetic flux contribution to
the BPS energy. The remaining three terms are total derivatives whose
integration under boundary conditions (\ref{bc00},\ref{bc01}) gives null
contribution to the total BPS energy. Thus,\
\begin{align}
\bar{E}_{BPS}& ={\pm }\int d^{2}t~{\bar{\mathcal{E}}\,}_{BPS}={\pm }\int
d^{2}t~\bar{B}\left( t\right)   \notag \\
&  \label{flux_ena} \\
& ={\pm 2\pi }\int_{0}^{\infty }dt~t\left( -\frac{\bar{a}^{\prime }}{t}%
\right) ={\pm }2\pi n.  \notag
\end{align}%
This shows that the\ BPS vortex solutions present energy or magnetic flux
quantization.\ Next, by using the BPS equations (\ref{eqt2}) and the Gauss's
law\ (\ref{eqt4}), the BPS energy density (\ref{ebps0})\ takes the suitable
form\
\begin{equation}
\bar{\mathcal{E}}_{BPS}=\left( 1-s\right) \bar{B}^{2}+2\frac{\bar{a}^{2}\bar{%
g}^{2}}{t^{2}}+{2\bar{g}^{2}\bar{\omega}}^{2}+\left( 1+\lambda _{r}\right)
\left( \bar{\omega}^{\prime }\right) ^{2},  \label{H_DEF_POS}
\end{equation}%
which is positive-definite for
\begin{equation}
s<1~~,~\ \ \lambda _{r}>-1.  \label{post1}
\end{equation}

\subsection{Asymptotic behavior}

The asymptotic behavior for $t\rightarrow 0$\ is obtained solving Eqs. (\ref%
{eqt2}-\ref{eqt4}) using power-series method. Thus, we\ attain
\begin{align}
& \displaystyle{\bar{g}\left( t\right) =Gt^{n}+\mathcal{\ldots }}
\label{bc0_g} \\[0.08in]
& \displaystyle{\bar{a}\left( t\right) =n-\frac{1}{2}{\frac{\left( 1+\lambda
_{r}\right) }{\left( 1-s\right) \left( 1+\lambda _{r}\right) +\kappa ^{2}}}%
t^{2}+\mathcal{\ldots }}  \label{bc0_a} \\[0.2cm]
& \displaystyle{\bar{\omega}\left( t\right) =\omega _{{0}}+{\frac{\kappa }{%
\left( 1-s\right) \left( 1+\lambda _{r}\right) +\kappa ^{2}}}t+\mathcal{\
\ldots }}  \label{bc0_w}
\end{align}%
From Eqs. (\ref{bc0_a}) and (\ref{cb1}), the magnetic field in the origin $%
\left( t=0\right) $\ is given by
\begin{equation}
\bar{B}\left( 0\right) =\frac{\left( 1+\lambda _{r}\right) }{\left(
1-s\right) \left( 1+\lambda _{r}\right) +\kappa ^{2}},  \label{b0}
\end{equation}%
while Eq. (\ref{bc0_w}) yields the electric field $\bar{\omega}^{\prime }$\
at $t=0$,
\begin{equation}
\bar{\omega}^{\prime }\left( 0\right) =\frac{\kappa }{\left( 1-s\right)
\left( 1+\lambda _{r}\right) +\kappa ^{2}},  \label{el0}
\end{equation}%
which establishes the second boundary condition for the field $\bar{\omega}%
(t)$. We should note that the denominator in the\ last two equations is
positive-definite due to the energy positivity conditions established in\
Eq. (\ref{post1}):\ $\left( 1-s\right) \left( 1+\lambda _{r}\right) +\kappa
^{2}>0$. Hence, the physical fields are well defined\ in the origin whenever
conditions (\ref{post1})\ are satisfied.

In the sequel we study the asymptotic behavior for $t\rightarrow +\infty $,
for which it holds $\bar{g}=1-\delta g_{1}$ , $\bar{a}=\pm \delta a_{1}$, $%
\bar{\omega}=\pm \delta \omega _{1}$, with $\delta g_{1}$, $\delta a_{1}$, $%
\delta \omega _{1}$\ being small corrections to be computed. After replacing
such forms in Eqs. (\ref{eqt2}-\ref{eqt4}), and solving the linearized set
of differential equations , we obtain
\begin{equation}
\delta g_{1}\sim t^{-1/2}e^{-\beta t}\sim \delta \omega _{1},\;\delta
a_{1}\sim t^{1/2}e^{-\beta t},
\end{equation}%
where $\beta $ is given as
\begin{equation}
\beta =\sqrt{\frac{2+\lambda _{r}-s\pm \sqrt{\left( \lambda _{r}+s\right)
^{2}-4\kappa ^{2}}}{\left( 1-s\right) \left( 1+\lambda _{r}\right) +\kappa
^{2}}}.  \label{beta}
\end{equation}%
Here, $(+)$ correspond to $\lambda _{r}+s>0$ and $(-)$ to $\lambda _{r}+s<0$%
, such that in the limit $\kappa =0$ we get the asymptotic behavior for the
BPS uncharged vortex, $\beta =\sqrt{2/(1-s)}$, see Ref. \cite{Carlisson1}.
We now analyze the $\beta -$parameter. First, the condition $2+\lambda
_{r}-s>0$ is guaranteed because of the energy positivity condition (\ref%
{post1}). The same holds for the denominator $\left( 1-s\right) \left(
1+\lambda _{r}\right) +\kappa ^{2}>0$. On the other hand, the term inside
the square root in the numerator, $\left( \lambda _{r}+s\right) ^{2}-4\kappa
^{2}$, is not definite-positive. Thus, we have a region $\left\vert \lambda
_{r}+s\right\vert \geq 2\kappa $ where $\beta $ is a positive real number,
yielding an exponentially decaying asymptotic behavior. On the other hand,
in the region $\left\vert \lambda _{r}+s\right\vert <2\kappa $ the parameter
$\beta $ becomes a complex number with positive real part, which implies a
sinusoidal behavior modulated by an exponentially decay factor.

\subsection{Numerical solutions}

We now analyze the case where $\beta $ is a real number by setting $\lambda
_{r}=0$ and $s=2\kappa $, so that $\beta =\sqrt{2/(1-\kappa )}$.
Consequently, the only free parameter is $\kappa ,$ whose values assuring a
positive-definite energy density are $\kappa <1/2-$ in accordance with the
condition (\ref{post1}).

\begin{figure}[H]
\begin{center}
\scalebox{1}[1]{\includegraphics[width=10cm,height=6cm]{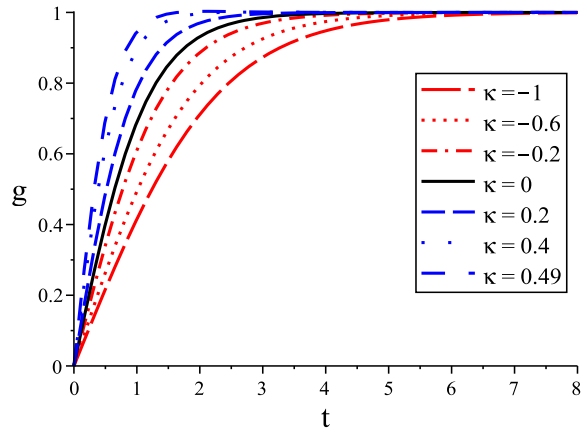}
}
\end{center}
\par
\vspace{-0.75cm}
\caption{Scalar field $\bar{g}(t)$ (Solid black line, $\protect\kappa =0$,
is the BPS solution for the Maxwell-Higgs model).}
\label{S_BPS}
\end{figure}

\begin{figure}[H]
\begin{center}
\scalebox{1}[1]{\includegraphics[width=10cm,height=6cm]{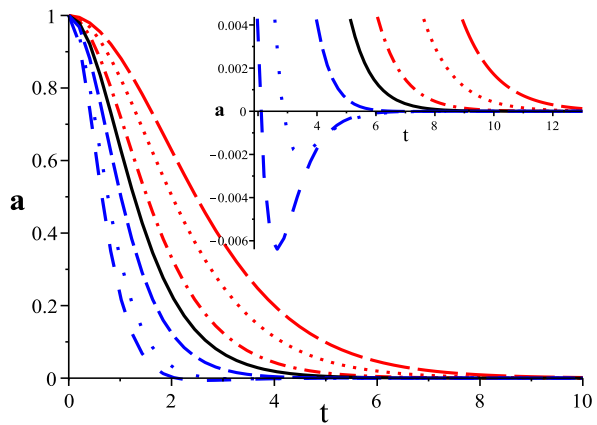}
}
\end{center}
\par
\vspace{-0.75cm}
\caption{Vector potential $\bar{a}(t)$.}
\label{A_BPS}
\end{figure}

\begin{figure}[H]
\begin{center}
\scalebox{1}[1]{\hspace{0.15cm}
\includegraphics[width=10cm,height=6cm]{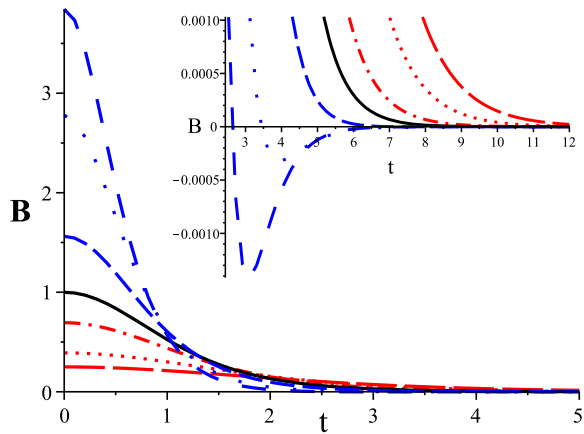}}
\end{center}
\par
\vspace{-0.75cm}
\caption{Magnetic field $\bar{B}(t)$.}
\label{B_BPS}
\end{figure}

\begin{figure}[H]
\begin{center}
\scalebox{1}[1]{\hspace{0.15cm}
\includegraphics[width=10cm,height=6cm]{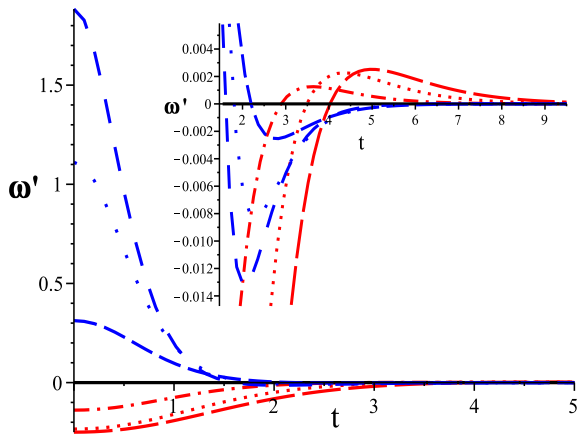}}
\end{center}
\par
\vspace{-0.5cm}
\caption{Electric field $\bar{\protect\omega}^{\prime }(t)$.}
\label{El_BPS}
\end{figure}

\begin{figure}[H]
\begin{center}
\scalebox{1}[1]{\includegraphics[width=10cm,height=6cm]{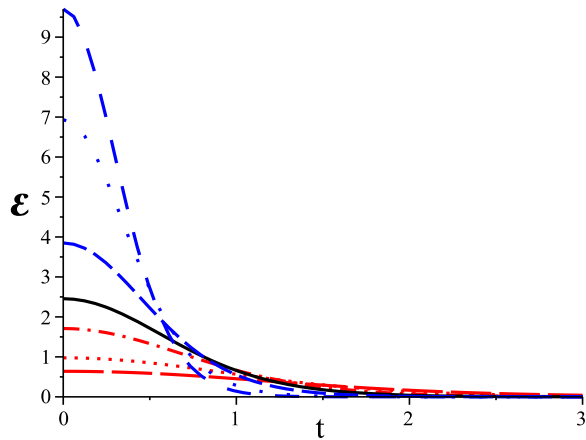}
}
\end{center}
\par
\vspace{-0.75cm}
\caption{Energy density $\bar{\mathcal{E}}(t)$.}
\label{Energy_BPS}
\end{figure}

In Figs. \ref{S_BPS}--\ref{Energy_BPS} we present some profiles (for the
winding number $n=1$) generated by numerical integration of the equations (%
\ref{eqt2}-\ref{eqt4}) using the Maple 13 libraries for solving\ the coupled
nonlinear differential equations. In all figures the value $\kappa =0$
reproduces the profile of the Maxwell-Higgs vortex \cite{ANO2} which is
depicted by a solid black line.\ The legends given in Fig. \ref{S_BPS} hold
for all figures.

Figs. \ref{S_BPS} and \ref{A_BPS} depict the numerical results obtained for
the Higgs field and vector potential, whose profiles are drawn around the
ones corresponding to the Maxwell-Higgs model. These profiles become wider
and wider for $\kappa <0$ and increasing $|\kappa |$, reaching more slowly
the respective saturation region. Otherwise, for $0<\kappa <1/2$ it
continuously shrinks approaching the minimum thickness when $\kappa \ $%
approaches to $1/2$. Moreover, the vector potential profile displays a
novelty: for $0<\kappa <1/2$\ it assumes negative values over a small region
of the radial axis (see insertion in Fig. \ref{A_BPS}), which is obviously
associated with a localized magnetic flux inversion. This inversion becomes
more pronounced for $\kappa $\ values near to $1/2$. Also, the region
presenting localized magnetic flux inversion is a little shifted to the
origin when $\kappa $\ tends to $1/2$.

Fig. \ref{B_BPS} depicts the magnetic field behavior. The profiles are lumps
centered at the origin whose amplitudes are proportional to $(1-\kappa )^{-2}
$. For $\kappa <0$ and increasing values of $|\kappa |$, the magnetic field
profile becomes increasingly wider with continuously diminishing amplitude.
For $0<\kappa <1/2$, the profile becomes narrower and higher for an
increasing $\kappa $, reaching its maximum value for $\kappa =1/2$. A close
zoom on the profiles corresponding to $\kappa \ $closer to $1/2$\ (see
insertion in Fig. \ref{B_BPS}) reveals that the magnetic field flips its
signal, showing explicitly localized magnetic flux reversion.

In a first view, Fig. \ref{B_BPS} seems to show that the magnetic flux
varies with the value of $\kappa $. Note, however, that the magnetic flux is
calculated by integrating $2\pi t\bar{B}(t)$, as showed in Eq. (\ref%
{flux_ena}). An explicit plot of the function $2\pi t\bar{B}(t)$ clearly
shows that for every $\kappa <1/2$ the magnetic flux is the same
independently of the localized magnetic field reversion. This result is in
accordance with the properties of the BPS-vortex solutions.

Fig. \ref{El_BPS} shows that electric field profiles also are lumps centered
at the origin with amplitudes proportional to $\kappa /(1-\kappa )^{2},$
having a minimum value for $\kappa =-1$\ and maximal value for $\kappa =1/2$%
. As it occurs with the magnetic field, the electric profiles become
narrower and higher while $\kappa $\ increases tending to $1/2$. Now, the
difference is that, for $\kappa <0,$ the profiles become negative (as
predicted after BPS equations). A close zoom along the $t-$axis (see
insertion in Fig. \ref{El_BPS}), however, reveals that the electric field
undergoes inversion both for positive and negative values of $\kappa $. Such
reversion is ubiquitous in all profiles.

The BPS energy density (\ref{H_DEF_POS}) is exhibited in Fig. \ref%
{Energy_BPS}, having profiles very similar to the magnetic field ones, being
more localized and possessing a higher amplitude, however. As the BPS\
energy density is positive-definite, no inversion regions are observed,
naturally.

By looking at the profiles of the BPS solutions it is observed that the
spatial thickness is controlled by the Lorentz-violating parameter, allowing
to obtain compactlike defects as in\ the uncharged case analyzed in Ref.
\cite{Carlisson1}. The present model is being regarded as an effective
electrodynamics, which subjected to the usual vortex ansatz, provides vortex
solutions in a dielectric continuum \cite{Dielectric}, as already mentioned
in Ref. \cite{Carlisson1}. This interpretation allows to consider
Lorentz-violating parameters with magnitude above the values stated by the
known vacuum upper-bounds.

These charged vortex configurations are endowed with several interesting
features, as space localization (exponentially decaying behavior), integer
magnetic flux quantization, magnetic flux and electric field reversion.
Specifically, the localized magnetic flux inversion is a very interesting
phenomenon, with sensitive appeal in condensed matter superconducting
systems. Recently, a magnetic inversion was reported in the context of
fractional vortices in superconductors described the two-component
Ginzburg-Landau (TCGL) model \cite{PRL1}. In it, the magnetic flux is
fractionally quantized and delocalized, presenting a $1/r^{4}$ decaying
behavior, and a subtle reversion. Such scenario, however, differs from the
one described by our theoretical model, which provides a localized magnetic
flux of controllable extent, exhibiting flux inversion just for the
parameters that yields narrower (compactlike) profiles.

One should still note that the set of parameters here analyzed is only one
of the many theoretical possibilities available. So, it can exist a set of
parameters for which the magnetic flux might undergo a more accentuated
reversion, reaching more appreciable flipping magnitudes. Such behavior
could be associated with a complex $\beta $ parameter, yielding oscillating
solutions which become less localized (and more similar with the ones of
Ref. \cite{PRL1}). Another interesting way is to investigate this
theoretical system in the presence of Lorentz-violating terms in the Higgs
sector. Such analysis are under progress with results being reported
elsewhere.

The authors are grateful to CNPq, CAPES and FAPEMA (Brazilian research
agencies) for invaluable financial support. CM, RC and MMFJr also
acknowledge the Instituto de F\'{\i}sica Te\'{o}rica (UNESP-S\~{a}o Paulo
State University) for the kind hospitality during the realization of part of
this work. E. da Hora thanks the Department of Mathematical Sciences of
Durham University (U.K.), for all their hospitality while doing part of this
work.

\end{document}